\documentclass[10pt]{article}
\usepackage{amsmath,amssymb,amsfonts,amsthm,graphicx}
\title{Quantum phase properties associated to solvable quantum systems using the nonlinear coherent states approach}
\author{G.R. Honarasa,
M. K. Tavassoly  and
M. Hatami
\\
\footnotesize{Atomic and Molecular Group, Faculty  of Physics, Yazd University, Yazd, Iran}
\\ \footnotesize{e-mail: mktavassoly@yazduni.ac.ir  } }

\begin{document}

\maketitle \thispagestyle{empty}
\pagestyle{headings} \markright{}

 \begin{abstract}
  In this paper we study the quantum phase properties of {\it "nonlinear coherent states"} and {\it "solvable quantum systems with discrete spectra"} using the Pegg-Barnett formalism in a unified approach. The presented procedure will then be applied to few special solvable quantum systems with known discrete spectrum as well as to some new classes of nonlinear oscillators with particular nonlinearity functions. Finally the associated phase distributions and their nonclasscial properties such as the squeezing in number and phase operators have been investigated, numerically.

 \end{abstract}

 {\bf keyword:}
   nonlinear coherent states, solvable quantum systems, phase distribution


% MATH -----------------------------------------------------------
\newcommand{\I}{\mathbb{I}}
\newcommand{\norm}[1]{\left\Vert#1\right\Vert}
\newcommand{\abs}[1]{\left\vert#1\right\vert}
\newcommand{\set}[1]{\left\{#1\right\}}
\newcommand{\R}{\mathbb R}
\newcommand{\C}{\mathbb C}
\newcommand{\DD}{\mathbb D}
\newcommand{\eps}{\varepsilon}
\newcommand{\To}{\longrightarrow}
\newcommand{\BX}{\mathbf{B}(X)}
\newcommand{\HH}{\mathfrak{H}}
\newcommand{\D}{\mathcal{D}}
\newcommand{\N}{\mathcal{N}}
\newcommand{\W}{\mathcal{W}}
\newcommand{\RR}{\mathcal{R}}
\newcommand{\HD}{\hat{\mathcal{H}}}

  %==========================================================================
 \section{Introduction}\label{sec-intro}
  %==========================================================================
    Coherent states defined as the right eigenstates of the harmonic oscillator annihilation operator, i.e.,  $a|z\rangle=z|z\rangle $ play an important role in quantum optics and modern physics \cite{ali1}. Along generalization of these states, nonlinear coherent states \cite{filho} or $f$-coherent states \cite{manko} have been introduced and attracted much attention in recent decade \cite{broy,sivakumar,sivakumar1}. According to this formalism two new operators $A$ and $A^\dag$ ($f$-deformed annihilation and creation operators, respectively) with an intensity dependent  function $f(n)$ defined as
    $A=af(n)=f(n+1)a$ and so $A^\dag=f^\dag(n)a^\dag=a^\dag f^\dag(n+1)$
   where $a$, $a^\dag $ and $n=a^\dag a$ are bosonic annihilation, creation and number operator, respectively \cite{manko}. Nonlinear coherent states are then defined as the right eigenstates of the generalized annihilation operator $A$, i.e.,
 \begin{equation}\label{eigenstate}
   A|z,f\rangle=z|z,f\rangle,
 \end{equation}
    where $z=\left|z\right|e^{i\phi}$. In the number state bases the explicit form of these states are given by
 \begin{equation}\label{newket}
   |z,f\rangle=\mathcal{N}_f(\left|z\right|^2)^{-1/2}\sum_{n=0}^\infty d_n z^n|n\rangle,
 \end{equation}
    with the coefficients $d_n$ as
 \begin{equation}\label{dn}
     d_n\doteq[\sqrt{n!}\prod_{i=1}^n f(i)]^{-1}
 \end{equation}
  and by definition $d_0\dot{=}1$. $\mathcal{N}_f(\left|z\right|^2)$ in (\ref{newket}) denotes a normalization constant and can be fixed up using $\langle z,f|z,f\rangle=1$, i.e.,
 \begin{equation}\label{N}
   \mathcal{N}_f(\left|z\right|^2)=\sum_{n=0}^\infty d_n ^2 \left|z\right|^{2n}.
 \end{equation}
    Note that we confine ourselves to the special case of real valued functions $f(n)$. With the help of equation (\ref{dn}) the function $f(n)$ corresponding to any set of generalized coherent states is found to be related to the expansion coefficients by
 \begin{equation}\label{fn}
     f(n)=\frac{1}{\sqrt{n}} \frac{d_{n-1}}{d_n}.
 \end{equation}
     Man'ko $et$ $al$ introduced the Hamiltonian of the deformed oscillator analogously to the harmonic oscillators as \cite{manko}
 \begin{equation}\label{manko}
      H_M=\frac{1}{2} (A^\dag A+AA^\dag).
 \end{equation}
    But later, Roknizadeh $et$ $al$ derived a Hamiltonian associated to a nonlinear system which was indeed the normal-ordered form of  Man'ko's Hamiltonian based on action identity requirement of nonlinear coherent states as follows \cite{roknizadeh}
  \begin{equation}\label{tavassoly}
     H=A^\dag A=nf^2(n).
  \end{equation}
    So the eigenvalue equation for any physical system with known discrete eigenvalues may be given by
  \begin{equation}\label{eigenvalue}
     H|n\rangle=e_n|n\rangle=nf^2(n)|n\rangle.
  \end{equation}
    Thus one simply has
  \begin{equation}\label{en}
    f(n)=\sqrt{e_n/n}.
  \end{equation}
  which indicates that the nonlinearity function corresponding to any quantum system with known discrete spectrum can be easily found. Therefore, the generalized coherent states associated with arbitrary solvable quantum system in terms of its spectrum may be introduced explicitly as
   \begin{equation}\label{newkete}
     |z,e_n\rangle=\mathcal{N}_e(\left|z\right|^2)^{-1/2}\sum_{n=0}^\infty {\frac {z^n}{\sqrt{[e_n]!}}|n\rangle},
   \end{equation}
  where
\begin{equation}\label{Ne}
   \mathcal{N}_e(\left|z\right|^2)=\sum_{n=0}^\infty \frac{\left|z\right|^{2n}}{[e_n]!}
\end{equation}
  and $[e_n]!\doteq e_n\; e_{n-1}\;...\;e_1$.
  Note that a necessary condition for the proposal in (\ref{tavassoly}) and (\ref{en}) is $e_0=0$ and so $[e_0]!=1$. It is straightforward to check that the following eigenvalue equation holds:  $A|z,e_n\rangle\equiv af(n)|z,e_n\rangle \equiv a\sqrt{\frac{e_n}{n}}|z,e_n\rangle = z |z,e_n\rangle$. It must be noted that the $\sqrt{\frac{e_n}{n}}$ factor is an operator-valued function which acts on the bases $\left\{|n\rangle\right\}_{n=0}^\infty$.

  On the other hand, the notion of quantum phase is an important concept in physics, especially in quantum optics. It has been given a great deal of attention for a long time, both theoretically \cite{dirac, susskind, barnet, pegg1,pegg, pegg2} and experimentally \cite{smithey,Ahmad,Mudassar}. Searching for a hermitian phase operator of radiation field has a long history from the beginnings of quantum electrodynamics \cite{dirac}. The initial attempts to construct explicitly a quantum phase operator as a quantity conjugate to the number operator were made by Dirac \cite{dirac}. The idea was to perform a polar decomposition of the annihilation operator, similar to the polar decomposition of the complex amplitude performed for classical fields. Susskind and Glogower \cite{susskind} propose an exponential phase operator based on Dirac's idea, with an extra condition related to vacuum. Unfortunately, although their formalism permits to define associated hermitian operators, it has still a serious problem: it is non-unitary. Later Barnett and Pegg \cite{barnet} introduced a unitary operator in an extended Hilbert space. The apparent difficulty of the latter proposal is that it included unphysical negative number states. Along these efforts, a very important development in this field has been made by Pegg and Barnett \cite{pegg1, pegg, pegg2}. They have defined a unitary and hermitian phase operator, and a phase state but in a finite although arbitrarily large subspace whose dimension was allowed to tend to infinity after the calculation of expectation values of observable quantities and moments \cite{pegg3} (the point that we concern with it when deal with uncertainties in phase and number operators). Up to now, the above three formalisms have been used along each other in the studies of phase properties and the phase fluctuations of various physical systems. A deep investigation in the three suggestions shows that all of them are problematic and the exact description of phase operator is still an open problem (for a review on the subject see \cite{shleich}). Altogether, in quantum optics the Pegg-Barnett formalism has been extensively employed in recent literature on the phase properties of a wide variety of quantum systems \cite{vaccaro and others}. We choose this formalism also due to the fact that it gives the ground state $|0\rangle$ a random phase which avoids some of the drawbacks in previous developments.  \\
   Phase properties of nonlinear coherent states associated to the center of mass motion of a trapped ion have been studied in the literature \cite{royy,xiang}. This is mainly due to the increasing interest in their nonclassical properties. We continued this formalism with a few new classes of nonlinear coherent states such as Penson- Solomon nonlinear coherent states and SU(1,1) coherent states in two distinct representation, i.e., Barut-Girardello and Gilmore-Perelomov. Nevertheless, to the best of our knowledge they are limited to specific systems with known nonlinearity functions. Generally, nonlinearity  functions correspond to some nonlinear oscillator algebras comes out from the algebraic deformations procedure. By this we mean that they are not directly extracted from a physical potential. The main goal of the present manuscript is to extend the approach to arbitrary quantum system with known discrete spectrum using a simple idea based on the nonlinear coherent states approach. Our motivation is to provide a general framework to investigate the phase distribution, and squeezing in phase and number operators as nonclassical features of  some quantum systems such as Hydrogen-like atoms, P\"{o}schl-Teller potentials and Isotonic oscillator.

  The paper is organized as follows. Keeping in mind the Pegg-Barnett phase distribution formalism for "nonlinear coherent states" we extend it to "solvable quantum systems" in the next section. There, we will obtain the phase distribution of both categories, i.e., in terms of $f(n)$ or $e_n$. Then in section 3 squeezing properties in phase or number operator corresponding to nonlinear coherent states and solvable quantum systems are studied in a general framework. Finally in section 4 the presented approach are applied to a few classes of generalized coherent stats with known nonlinearity functions as well as to some solvable quantum systems and the nonclassicality of the obtained  states are investigated numerically.

 %============================================================================================
 \section{Phase distribution of solvable quantum systems}
         \label{sec-nl}
%================================================================================================
 According to the Pegg-Barnett formalism a complete set of $s+1$ orthonormal phase states $|\theta_P\rangle$ are defined by \cite{pegg}
  \begin{equation}\label{tetaket}
   |\theta_P\rangle=\frac{1}{\sqrt{s+1}}\sum_{n=0}^s \exp{(in\theta_P)} |n\rangle,
 \end{equation}
   where $\left\{|n\rangle\right\}_{n=0}^s$ are the number states and $\theta_P$ is given by
 \begin{equation}\label{teta}
   \theta_P=\theta_0 +\frac{2\pi P}{s+1} ,\qquad   P=0,1,...,s,
 \end{equation}
    with arbitrary $\theta_0$ value. Based on the phase states definition in (\ref{tetaket}) the Hermitian phase operator is defined as
 \begin{equation}\label{phi}
     \phi_\theta=\sum_{P=0}^s \theta_P |\theta_P\rangle \langle \theta_P|.
 \end{equation}
    The Pegg-Barnett phase probability distribution of nonlinear coherent states may be defined as \cite{roy}
 \begin{equation}\label{p}
     P_f(\theta)=\lim_{s\rightarrow\infty}\frac{s+1}{2\pi} \left|\langle\theta_m|z,f\rangle\right|^2.
 \end{equation}
    Using (\ref{tetaket}) we can also express (\ref{newket}) in the phase space bases as
 \begin{equation}\label{zf}
    |z,f\rangle=\frac{\mathcal{N}_f(|z|^2)^{-1/2}}{\sqrt{s+1}} \sum_{n} \sum_{m=0}^s e^{-in\theta_m} d_n z^n |\theta_m\rangle.
 \end{equation}
    Noticing that
 \begin{equation}\label{teta-z}
    \langle\theta_m|z,f\rangle=\frac{\mathcal{N}_f(|z|^2)^{-1/2}}{\sqrt{s+1}} \sum_{n} e^{-in\theta_m} d_n z^n,
 \end{equation}
   the phase distribution in (\ref{p}) for the nonlinear coherent states will be obtained as
 \begin{eqnarray}\label{pteta1}
    P_f(\theta)=\frac{\mathcal{N}(|z|^2)^{-1}}{2\pi}\left|\sum_{n} e^{-in\theta_n} d_n z^n\right|^2 \;\;\;\;\;\;\;\;\;\;\;\;\;\;\;\;\;\;\;\;\;\;\;\;\;\;\;\;\;\;\;\;\;\;\;\;\;\;\;\;\;\;\;\;\;\;\;\;\;\;\;\;\;\;\;\;\;\;\;\;\; \\
  =\frac{\mathcal{N}(|z|^2)^{-1}}{2\pi}\sum_{n} d_n^2 |z|^{2n} + \frac{2\mathcal{N}(|z|^2)^{-1}}{2\pi}\sum_{n} \sum_{k<n} d_n d_k z^n z^{*k} \cos{[(n-k)\theta]},\nonumber
 \end{eqnarray}
   where the terms corresponding to $n=k$ and $n\neq k$ were written separately. Thus, with the help of (\ref{N}) we will finally arrive at
 \begin{equation}\label{pteta3}
  P_f(\theta)=\frac{1}{2\pi}\left(1+2\mathcal{N}(|z|^2)^{-1}\sum_{n} \sum_{k<n} d_n d_k z^n z^{*k} \cos{[(n-k)\theta]}\right),
 \end{equation}
   with $d_n$ has been defined in (\ref{dn}).\\
   Following the same procedure lead us to (\ref{pteta3}) for nonlinear coherent states, we may obtain the generalized coherent states associated to solvable quantum systems and their phase distributions by introducing
 \begin{equation}\label{pe}
    P_e(\theta)=\lim_{s\rightarrow\infty}\frac{s+1}{2\pi} \left|\langle\theta_m|z,e_n\rangle\right|^2.
 \end{equation}
    Using (\ref{tetaket}) we can also express (\ref{newkete}) in the phase space bases as
 \begin{equation}\label{ze}
    |z,e_n\rangle=\frac{\mathcal{N}_e(|z|^2)^{-1/2}}{\sqrt{s+1}} \sum_{n} \sum_{m=0}^s \frac{z^n}{\sqrt{[e_n]!}}e^{-in\theta_m} |\theta_m\rangle.
 \end{equation}
   At last, the Pegg-Barnett phase probability distribution can be easily expressed for any quantum system in terms of its spectrum $e_n$ as
 \begin{equation}\label{pteta4}
    P_e(\theta)=\frac{1}{2\pi}\left(1+2\mathcal{N}_e(|z|^2)^{-1}\sum_{n} \sum_{k<n} {\frac{z^n z^{*k}}{\sqrt{[e_n]![e_k]!}}}\cos{[(n-k)\theta]}\right).
 \end{equation}
   To the best of our knowledge, there is no report on using Pegg-Barnett formalism for generalized coherent states associated to quantum systems with known discrete spectrum.
 %============================================================================================
 \section{Squeezing properties of solvable quantum systems}
         \label{sec-n2}
%================================================================================================

    Since $n$ and $\phi_\theta$ are conjugate operators they satisfy the uncertainty relation
 \begin{equation}\label{deltandeltaphi}
    \left\langle(\Delta\phi_\theta)^2\right\rangle \left\langle(\Delta n)^2\right\rangle \geq \frac {1}{4} |\langle [n,\phi_\theta] \rangle|^2,
 \end{equation}
   where it is proved that \cite{pegg}
 \begin{equation}\label{nphi}
    [n,\phi_\theta]=i[1-2\pi P(\theta_0)].
 \end{equation}
    If $\langle (\Delta n)^2\rangle<\frac {1}{2} |\langle [n,\phi_\theta] \rangle|$ or $\langle (\Delta\phi_\theta)^2\rangle<\frac {1}{2} |\langle [n,\phi_\theta] \rangle|$ the squeezing occurs in number or phase operator, respectively.\\
    The phase and number variances for nonlinear coherent states are given respectively by
  \begin{eqnarray}\label{deltaphi}
    \langle (\Delta\phi_\theta)^2\rangle_f=\int{\theta^2P_f(\theta)d\theta}-\left(\int{\theta P_f(\theta)d\theta} \right)^2 \qquad  \;\;\;\;\;\;\;\;\;\;\; \nonumber \\
    \qquad \;\;\;\;\;\, \;=\frac{\pi^2}{3}+4\mathcal{N}_f(|z|^2)^{-1}\sum_{n} \sum_{k<n} d_n d_k z^n z^{*k} \frac{(-1)^{n-k}}{(n-k)^2}
  \end{eqnarray}
  and
 \begin{eqnarray}\label{deltan}
    \langle (\Delta n)^2\rangle_f=\langle n^2\rangle- \langle n\rangle ^2 \qquad \qquad \;\;\;\;\;\;\;\;\;\;\;\;\;\;\;\;\;\;\;\;\;\;\;\;\;\;\;\;\;\;\;\;\;\;\;\;\;\;\;\;\;\;\;\;\;\;\;\;\;\;\;\;\; \nonumber \\
    \qquad \;\;\;\;\;\,=\mathcal{N}_f(|z|^2)^{-1}\sum_{n} n^2 d_n^2 |z|^{2n}-\left(\mathcal{N}_f(|z|^2)^{-1}\sum_{n} n d_n^2 |z|^{2n}\right)^2,
 \end{eqnarray}
    where in (\ref{deltaphi}) and (\ref{deltan}) we have used (\ref{pteta3}) and (\ref{newket}) respectively.\\
    Similarly, demanding for the phase and number variances corresponding to solvable physical systems one may obtain
 \begin{equation}\label{deltaphi2}
   \langle (\Delta\phi_\theta)^2\rangle_e=\frac{\pi^2}{3}+4\mathcal{N}_e(|z|^2)^{-1}\sum_{n} \sum_{k<n} {\frac{z^n z^{*k}(-1)^{n-k}}{\sqrt{[e_n]![e_k]!}}(n-k)^2}
 \end{equation}
   and
 \begin{equation}\label{deltan2}
  \langle (\Delta n)^2\rangle_e=\mathcal{N}_e(|z|^2)^{-1}\sum_{n} n^2 {\frac{|z|^{2n}}{[e_n]!}}-\left(\mathcal{N}_e(|z|^2)^{-1}\sum_{n} n{\frac{|z|^{2n}}{[e_n]!}}\right)^2
 \end{equation}
   where in (\ref{deltaphi2}) and (\ref{deltan2}) we have used (\ref{pteta4}) and (\ref{newkete}) respectively.\\
   To study nonclassical properties (the squeezing in number or phase components) the following squeezing parameters were introduced
 \begin{equation}\label{sn}
    S_n^{f(e)}=\frac{2\langle (\Delta n)^2\rangle_{f(e)}}{|\langle [n,\phi_\theta] \rangle_{f(e)}|} -1,
 \end{equation}
 \begin{equation}\label{sphi}
    S_\phi^{f(e)}=\frac{2\langle (\Delta\phi_\theta)^2\rangle_{f(e)}}{|\langle [n,\phi_\theta] \rangle_{f(e)}|} -1.
  \end{equation}
  If $S_n^{f(e)} <0$ ($S_\phi^{f(e)} <0$) then the associated state is number (phase) squeezed.

  %++++++++++++++++++++++++++++++++++============================================
  \section{Some physical realizations of the formalism and their nonclassical properties }
  %==============================================================================
   Now we are ready to apply the presented formalism to a few generalized coherent states with known nonlinearity functions $f(n)$ as well as quantum systems with known spectra $e_n$. For the first case we will deal with nonlinear coherent states introduced by Penson- Solomon \cite{penson-solomon} and SU(1,1) coherent states in two distinct representation, i.e., Barut-Girardello (BG) and Gilmore-Perelomov (GP) \cite{barut,Perelomov}. While for solvable quantum systems P\"{o}schl-Teller (PT) potential, infinite well, isotonic oscillator (IO) and Hydrogen-like atoms will be considered.
   %%%%%%%%%%%%%%%%%%%%%%%%%%%%%%%%%%%%%%%%%%%%%%%%%%%%%%%%%%%%%%%%%%%%%%%%%%%%%5
   %%%%%%%%%%%%%%%%%%%%%%%%%%%%%%%%%%%%%%%%%%%%%%%%%%%%%%%%%%%%%%%%%%%%%%
 \subsection{Nonlinear coherent states}
  %%%%%%%%%%%%%%%%%%%%%%%%%%%%%%%%%%%%%%%%%%%%%%%%%%%%%%%%%%%%%%%%%%%%%%%%%%%%%%%%
  In this subsection we will study the phase properties and squeezing parameters of three classes of nonlinear coherent states. Obviously it seems to be enough to introduce only the explicit form of nonlinearity function $f(n)$, since knowing it is capable to perform $P_f(\theta)$ by (\ref{pteta3}) and $S_n^{f}$, $S_\phi^{f}$ using (\ref{deltaphi}), (\ref{deltan}), (\ref{sn}) and (\ref{sphi}).
  %%%%%%%%%%%%%%%%%%%%%%%%%%%%%%%%%%%%%%%%%%%%%%%%%%%%%%%%%%%%%%%%%%%%%%%%
 \subsubsection{Penson-Solomon coherent states.}
%%%%%%%%%%%%%%%%%%%%%%%%%%%%%%%%%%%%%%%%%%%%%%%%%%%%%%%%%%%%%%%%%%%%%
    PS coherent states introduced as \cite{penson-solomon}:
 \begin{equation}\label{penson}
   |q,z\rangle_{PS}=\mathcal{N}_f(q,|z|^2)^{-1/2}\sum_{n=0}^\infty \frac{q^{n(n-1)/2}}{\sqrt{n!}}z^n|n\rangle,\;\;\;\;\;\;z\in \C,
 \end{equation}
   where $\mathcal{N}_f(q,|z|^2)^{-1/2}$ is an appropriate normalization function and $0\leq q \leq1$. The nonlinearity function can be easily obtained by (\ref{fn}) as
 \begin{equation}\label{fpenson}
     f_{PS}(n)=q^{1-n}.
 \end{equation}
   In figure 1 we have plotted the Pegg-Barnett phase distribution against $\theta$ for various values of $z\in \R$ using (\ref{pteta3}), keeping $q$ fixed at 0.5. From the figure it is seen that $P_f(\theta)$ has a single peak at $\theta=0$ and as $z$ increases the peak becomes sharper and higher. To observe squeezing behavior  in number or phase operators for PS coherent states we have plotted $S_n^f$ and $S_\phi^f$ against $z$ for $q=0.5$ in figure 2. From the figure we find that the squeezing parameters have opposite behavior, i.e., as $z$ increases $S_n^f$ increases while $S_\phi^f$ decreases. Also, for $z\lesssim2.4$, $S_n^f<0$ and $S_\phi^f>0$ implying squeezing occurance in the number component. For $2.4\lesssim z\lesssim2.7$ there is not any squeezing effect in number and phase components. But as $z$ increases from 2.7 to the larger values we find that $S_\phi^f$ becomes negative indicating squeezing in the phase operator.

  \subsubsection{Barut-Girardello coherent states of $SU(1,1)$ group.}
%%%%%%%%%%%%%%%%%%%%%%%%%%%%%%%%%%%%%%%%%%%%%%%%%%%%%%%%%%%%%%%%%%%%%%%%
   BG coherent states defined for the discrete series representations of the $SU(1,1)$ group \cite{barut}. These states decomposed over the number states bases as
  \begin{equation}\label{suBG}
   |z,\kappa\rangle^{su(1,1)}_{BG} = \mathcal{N}_f(|z|^2)^{-1/2}\sum_{n=0}^\infty
   \frac{z^n}{\sqrt{n!\Gamma(n+2\kappa)}} |n\rangle,\;\;\;\;\;\;z\in \C,
 \end{equation}
   where $\mathcal{N}_f(|z|^2)$  is the normalization constant and the label
    $\kappa$ takes the discrete values $1/2, 1, 3/2, 2,
    \cdots$.
    Using (\ref{fn}) the nonlinearity function in this case is determined as
     \begin{equation}\label{fsuBG}
    f_{BG}(n,\kappa)=\sqrt{n+2\kappa -1}.
  \end{equation}
    In figure 3 we have displayed the Pegg-Barnett phase distribution for BG coherent states  against $\theta$ for different values of $z\in \R$, keeping $\kappa$ fixed at 3. The behavior of $P_f(\theta)$ in this case is qualitatively similar to figure 1. From figure 4 we find that the squeezing parameters have opposite behaviors and show number and phase squeezing in small and large values of $z$, respectively. For the interval $2.05\lesssim z\lesssim2.75$ squeezing effect neither in number nor in phase components is observed.

 %%%%%%%%%%%%%%%%%%%%%%%%%%%%%%%%%%%%%%%%%%%%%%%%%%%%%%%%%%%%%%%%%
 \subsubsection{Gilmore-Perelomov coherent states of $SU(1,1)$ group.}
 %%%%%%%%%%%%%%%%%%%%%%%%%%%%%%%%%%%%%%%%%%%%%%%%%%%%%%%%%%%%%%%%%%%%
   Next we deal with GP coherent states of $SU(1,1)$ group
   whose the number state representation read as \cite{Perelomov}:
  \begin{equation}\label{suGP}
   |z, \kappa\rangle^{su(1,1)}_{GP} = \mathcal{N}_f(|z|^2)^{-1/2}\sum_{n=0}^\infty
   \sqrt{\frac{\Gamma(n+2\kappa)}{n!}}z^n \; |n\rangle,\;\;\;\;\;\;|z|<1,
  \end{equation}
    where $\mathcal{N}_f(|z|^2)$  is the normalization constant and the label
    $\kappa$ takes the discrete values $1/2, 1, 3/2, 2,
    \cdots$.
    The  nonlinearity function in this case may be determined by (\ref{fn}) as
  \begin{equation}\label{fsu}
    f_{GP}(n,\kappa)=\frac{1}{\sqrt{n+2\kappa -1}}.
  \end{equation}
   Recall that the states in (\ref{suBG}) and (\ref{suGP}) have been known as the dual pair family \cite{roknizadeh,ali}. In figure 5 we have plotted the Pegg-Barnett phase distribution for GP coherent states against $\theta$ for various values of $z\in \R$, keeping $\kappa$ fixed at 3. From the figure it is seen that $P_f(\theta)$ has a single peak at $\theta=0$ and as $z$ increases becomes sharper and higher, similar to the cases were happened in figure 1 and 3. We have plotted the squeezing parameters against $z$ for GP coherent states in figure 6. From the figure we find that for $z\lesssim0.29$ number squeezing occurs. One also observe that in the interval $0.29\lesssim z\lesssim0.63$ both $S_n^f$ and $S_\phi^f$ are positive, which implies the absence of squeezing in both components in the mentioned region. Also, for $z\gtrsim0.63$  $S_\phi^f$ is negative indicating squeezing in the phase component.

%%%%%%%%%%%%%%%%%%%%%%%%%%%%%%%%%%%%%%%%%%%%%%%%%%%%%%%%%%5
\subsection{Quantum systems with known discrete spectra}
%%%%%%%%%%%%%%%%%%%%%%%%%%%%%%%%%%%%%%%%%%%%%%%%%%%%%%%%%%%%%%%%%
  In this subsection we applied the presented formalism in the paper to Hydrogen-like spectrum, P\"{o}schl-Teller potential and isotonic oscillator as samples of solvable quantum systems with discrete spectra. We will only introduce the associated spectrum of any quantum system $e_n$ which will be considered and then the numerical results of $P_e(\theta)$ is given by (\ref{pteta4}) and $S_n^e$, $S_\phi^e$ will be expressed using (\ref{deltaphi2}), (\ref{deltan2}), (\ref{sn}) and (\ref{sphi}).
%%%%%%%%%%%%%%%%%%%%%%%%%%%%%%%%%%%%%%%%%%%%%%%%%%%%%%%%%%%%

 %%%%%%%%%%%%%%%%%%%%%%%%%%%%%%%%%%%%%%%%%%%
   \subsubsection{Hydrogen-like spectrum.}
 %%%%%%%%%%%%%%%%%%%%%%%%%%%%%%%%%%%%%%%%%%%%%%%%%%%%%%%%%%%%%%%%%
   As an important physical system we will accomplish in the present paper
   we want to apply our proposal into the
   hydrogen-like spectrum. This quantum system is described by \cite{roknizadeh,tavassoly}
 \begin{equation}\label{H-spect}
   e_n= 1- \frac  {1}{(n+1)^2}\; .
  \end{equation}
   The associated coherent states may be obtained explicitly by setting (\ref{H-spect}) in (\ref{newkete}) which are defined in the unit disk centered at the origin.
   In figures 7 and 8 we have plotted the Pegg-Barnett phase distribution and the squeezing parameters for Hydrogen-like spectrum, respectively. From figure 7 it is seen that $P_e(\theta)$ has a single peak at $\theta=0$ and as $z$ increases the peak becomes sharper and higher. From figure 8 we find that the squeezing parameters have opposite behavior. In the range $z\lesssim0.42$, $S_n^e<0$ and $S_\phi^e>0$ implying squeezing occurance in the number component in this interval. For $0.42\lesssim z\lesssim0.78$ no squeezing effect may be observed in number and phase components. For $z\gtrsim0.78$ we find that $S_\phi^e$ becomes negative, indicating squeezing in the phase operator.

 %%%%%%%%%%%%%%%%%%%%%%%%%%%%%%%%%%%%%%%%%%%%%%%
  \subsubsection{P\"{o}schl-Teller potential.}
  %%%%%%%%%%%%%%%%%%%%%%%%%%%%%%%%%%%%%%%%%%%%%%
    The interest in this potential and its coherent states is due to various application in many fields of physics particularly in atomic and molecular physics. The explicit form of PT potential is as bellow
 \begin{equation}\label{poschlpotential}
    V(x)= \frac{1}{2}V_0\left(\frac{\lambda(\lambda-1)}{\cos^2\frac{x}{2a}}+\frac{\kappa(\kappa-1)}{\sin^2\frac{x}{2a}}\right),\qquad     0\leq x \leq \pi a,
  \end{equation}
    where $\lambda,\kappa>1$ and $V_0$ is a coupling constant. The following spectrum is known for this potential \cite{antoine}
  \begin{equation}\label{poschl}
   e_n= n(n+\nu),
  \end{equation}
    where $\nu=\lambda+\kappa>2$.
    The spectrum of the infinite square-well potential may be obtained by setting $\nu=2$ in equation (\ref{poschl}). The associated coherent states may be obtained explicitly by setting (\ref{poschl}) in (\ref{newkete}) which are defined in the whole of complex plane.\\
    In figures 9 and 10 we have plotted the Pegg-Barnett phase distribution and the squeezing parameters for the PT potential, keeping $\nu$ fixed at 5, respectively. The behavior of $P_e(\theta)$ and the squeezing parameters for PT potential are qualitatively similar to the Hydrogen-like spectrum except that the distance with no squeezing change to $1\lesssim z\lesssim2.84$. We have also done the numerical calculation for infinite well potential setting $\nu=2$. No qualitative difference has been observed and both of them exhibit nonclassical properties through showing squeezing effects in number or phase component in the some regions.

 %%%%%%%%%%%%%%%%%%%%%%%%%%%%%%%%%%%%%%%%%%%%%%%%%%%
 \subsubsection{Isotonic oscillator.}
 %%%%%%%%%%%%%%%%%%%%%%%%%%%%%%%%%%%%%%%%%%%%%%%
    An interesting model of a solvable system is the isotonic oscillator Hamiltonian \cite{landau}
 \begin{equation}\label{isotonic}
   H= -\frac{d^2}{dx^2}+x^2+\frac{A}{x^2} ,\qquad    A\geq0 .
     \end{equation}
    It is known that the Hamiltonian in (\ref{isotonic}) admits exact eigenvalues given by \cite{richard,saad}
 \begin{equation}\label{isot}
   \epsilon_n= 2(2n+\gamma) , \qquad       n=0,1,2,...,
 \end{equation}
    where $\gamma=1+\frac{1}{2}\sqrt{1+4A}$. For satisfying the condition $e_0=0$ we have to use the shifted energy
  \begin{equation}\label{isot2}
   e_n= \epsilon_n-2\gamma=4n , \qquad       n=0,1,2,... .
 \end{equation}
   The associated coherent states may be obtained explicitly by setting (\ref{isot2}) in (\ref{newkete}) which are defined in the whole of the complex plane.
   In figures 11 and 12 we have plotted the Pegg-Barnett phase distribution and the squeezing parameters for the isotonic oscillator, keeping $\gamma$ fixed at 5/2. The behavior of $P_e(\theta)$ is qualitatively similar to figure 7 and figure 12  shows number and phase squeezing in small and large values of $z$, respectively. For the interval $1.2\lesssim z\lesssim2.2$, squeezing effect neither in number nor in phase component is observed.
  %=========================================================================================
    \section{Summary and conclusion}
 %============================================================================================
   Upon much interests in the phase and number properties of physical systems both theoretically \cite{royy, xiang} and experimentally \cite {Ahmad, Mudassar, Faisal}, in summary we have shown that the Pegg-Barmnett formalism of nonlinear coherent states can be extended to arbitrary solvable quantum systems with discrete spectra. We have then examined the phase properties of nonlinear coherent states as well as the generalized coherent states associated to solvable quantum systems in a general framework.
   To this end, using the phase space expansions of the corresponding coherent states of "nonlinear coherent states" and "solvable quantum systems", {\it the phase probability distribution} and also {\it phase} and {\it number squeezing} of a few physical systems with known nonlinearity function $f(n)$ or discrete spectrum $e_n$ have been investigated, numerically. It is found that these states exhibit phase or number squeezing for different ranges of real values of $z$. Consequently all of the introduced states possess nonclassical signatures.
   Adding our results we have demonstrated that how a simple idea can provide a powerful method that enables one to study the phase properties of real physical systems with particular potential.
   It is then a straightforward matter to apply the presented formalism to other physical systems as photon added (and subtracted) coherent states \cite{sivakumar}, pseudoharominc oscillator \cite{popov}, ...  also their dual family \cite{ali, tavassoly}.

%==============================================
\newpage
%==============================================
{\bf FIGURE CAPTIONS}
%==============================================
  \vspace {1.5 cm}

  {\bf FIG. 1} Pegg-Barnett phase distribution against $\theta$ for various values of $z\in \R$ associated to PS coherent states.

  \vspace {.5 cm}

    {\bf FIG. 2} Plot of $S_n^f$ (the dashed curve) and $S_\phi^f$ (the solid curve) against $z\in \R$ associated to PS coherent states.
  \vspace {.5 cm}

    {\bf FIG. 3.} Pegg-Barnett phase distribution against $\theta$ for various values of $z\in \R$ associated to BG coherent states.
  \vspace {.5 cm}

    {\bf FIG. 4} Plot of $S_n^f$ (the dashed curve) and $S_\phi^f$ (the solid curve) against $z\in \R$ associated to BG coherent states.
  \vspace {.5 cm}

  {\bf FIG. 5}
   Pegg-Barnett phase distribution against $\theta$ for various values of $z\in \R$ associated to GP coherent states.
  \vspace {.5 cm}

  {\bf FIG. 6} Plot of $S_n^f$ (the dashed curve) and $S_\phi^f$ (the solid curve) against $z\in \R$ associated to GP coherent states.

  \vspace {.5 cm}

    {\bf FIG. 7} Pegg-Barnett phase distribution against $\theta$ for various values of $z\in \R$ associated to hydrogen-like spectrum.
  \vspace {.5 cm}

    {\bf FIG. 8}  Plot of $S_n^e$ (the dashed curve) and $S_\phi^e$ (the solid curve) against $z\in \R$ associated to hydrogen-like spectrum.

  \vspace {.5 cm}

    {\bf FIG. 9}  Pegg-Barnett phase distribution against $\theta$ for various values of $z\in \R$ associated to PT potential.
  \vspace {.5 cm}

    {\bf FIG. 10}  Plot of $S_n^e$ (the dashed curve) and $S_\phi^e$ (the solid curve) against $z\in \R$ associated to PT potential.

    \vspace {.5 cm}

    {\bf FIG. 11}  Pegg-Barnett phase distribution against $\theta$ for various values of $z\in \R$ associated to isotonic oscillator.

    \vspace {.5 cm}

    {\bf FIG. 12}  Plot of $S_n^e$ (the dashed curve) and $S_\phi^e$ (the solid curve) against $z\in \R$ associated to isotonic oscillator.

%==============================================================================

\end{document}